\documentclass{ws-procs9x6}

\begin{document}

\title{PHASE STATES AND COHERENT STATES FOR \\ 
       GENERALIZED WEYL-HEISENBERG ALGEBRAS$^\dagger$}

\author{M. R. KIBLER$^*$ and M. DAOUD}

\address{Institut de Physique Nucl\'eaire, Universit\'e de Lyon (Lyon 1),\\
Villeurbanne, F-69622, France\\
$^\dagger$Version: 14 October 2012 -- $^*$E-mail: kibler@ipnl.in2p3.fr}

\begin{abstract}
This report constitutes an introduction to three papers published by the authors in J. Phys. A 
[{\bf 43} (2010) 115303 and {\bf 45} (2012) 244036] and J. Math. Phys. [{\bf 52} (2011) 082101]. See 
these three papers for the relevant references. 
\end{abstract}


\bodymatter

\section{General outline}

This paper is concerned with the construction of phase operators, phase states, vector phase states, and coherent states for a generalized Weyl-Heisenberg algebra. This polynomial algebra (that depends on real parameters) is briefly described. The various states are defined on a finite- or infinite-dimensional space depending on the parameters.

\section{Generalized Weyl-Heisenberg Algebra}

\subsection{The ${\cal A}_{ \{ \kappa \} }(1)$ algebra}

Let ${\cal A}_{ \{ \kappa \} }(1)$ be the generalized Weyl-Heisenberg algebra spanned by $a^-$, $a^+$ and $N$ satisfying 
		\begin{eqnarray}
& & [a^- , a^+] = F(N+I) - F(N), \quad [N, a^{\pm}] = \pm a^{\pm} 
\nonumber 
\\
& & N = N^{\dagger}, \quad a^+ = (a^-)^{\dagger}, \quad 
F(N) = N \prod_{i = 1}^{r} [I + \kappa_i(N-I)], \quad \kappa_i \in \mathbb{R}.
\nonumber 
		\end{eqnarray}
This definition yields a multi-parameter 
polynomial Weyl-Heisenberg algebra depending on 
$r$ parameters  
$\{ \kappa \} \equiv \{ \kappa_1, \kappa_2, \ldots, \kappa_r \}$. The 
harmonic oscillator algebra corresponds to $\kappa_i = 0$
($F(N) = N \Rightarrow F(N+I) - F(N) = I$). 

\subsection{Hilbertian representation of ${\cal A}_{ \{\kappa\} }(1)$}

We suppose that the operators $a^{\pm}$ and $N$ act on a (finite or not) Hilbert space ${\cal F}_{\kappa}$ with an orthonormal basis 
$\{ \vert n \rangle : n \ {\rm ranging} \}$ arising from the eigenvectors of $N$. The following actions 
	  \begin{eqnarray} 
& & N   \vert n \rangle = n \vert n \rangle, \quad a^+ \vert n \rangle = \sqrt{F(n+1)} e^{{-i [F(n+1)- F(n)]  \varphi }} \vert n+1 \rangle 
		\nonumber  
\\ 
& & a^- \vert n \rangle = \sqrt{F(n)}   e^{{+i [F(n) - F(n-1)] \varphi }} \vert n-1 \rangle, \quad a^- \vert 0 \rangle = 0 
		\nonumber 
	  \end{eqnarray} 
(where $\varphi \in \mathbb{R}$) define an Hilbertian representation of ${\cal A}_{ \{\kappa\} }(1)$. 

The significance of the $F$ structure function is clear. The operator  
		\[
H = a^+ a^- = F(N) 
		\]
can be considered as the Hamiltonian for a quantum system. 

\subsection{Dimension of ${\cal F}_{\kappa}$}

The dimension of the ${\cal F}_{\kappa}$ representation space of ${\cal A}_{\{\kappa\}}(1)$ is controlled by 
$F(n) \geq 0$. Two interesting cases deserve to be considered: (i) $\kappa_i \geq 0$ ($i= 1, 2, \cdots, r$) $\Rightarrow$ 
infinite-dimensional representation and (ii) $\kappa_1 < 0$, $\kappa_i \geq 0$ ($i= 2, 3, \cdots, r$) $\Rightarrow$ 
finite representation of dimension 
		\begin{eqnarray}
d = 1 - \frac{1}{\kappa_1} \quad {\rm for} \quad -\frac{1}{\kappa_1} \in \mathbb{N}^*.
		\nonumber  
		\end{eqnarray}
The Fock-Hilbert space is spanned by $\{ | n \rangle : n \in \mathbb{N} \}$ in the 
infinite-dimensional case and by $\{ | n \rangle : n = 0, 1, \cdots, d-1 \}$ in the 
finite-dimensional case. In finite dimension, two further conditions are fulfilled, 
viz, $a^+ \vert d-1 \rangle = 0$ and $(a^-)^{d} = (a^+)^{d} = 0$, 
which generalize the conditions for $k$-fermions which are objects, introduced by 
the authors, interpolating between fermions ($k=d=2$) and bosons ($k=d \to \infty$).

\subsection{Example: The ${\cal A}_{ \kappa }(1)$ algebra}

This example corresponds to $r = 1$, $\kappa_1 = \kappa$, and ${\cal A}_{ \{\kappa\} }(1) \equiv {\cal A}_{ \kappa }(1)$. The 
${\cal A}_{ \kappa }(1)$ one-parameter algebra is then defined by 
      \begin{eqnarray}
[a^- , a^+] = I + 2 \kappa N,       \quad 
[N , a^{\pm}] = \pm a^{\pm},        \quad 
\left( a^- \right)^{\dagger} = a^+, \quad
N^{\dagger} = N                
      \nonumber 
      \end{eqnarray}
and gives rise to three cases: 
(i)   $\kappa = 0 \rightarrow$ ${\cal A}_{\kappa}(1) = h_4$ (usual Weyl-Heisenberg algebra); 
(ii)  $\kappa > 0 \rightarrow$ ${\cal A}_{\kappa}(1) = su_{1,1}$; and 
(iii) $\kappa < 0 \rightarrow$ ${\cal A}_{\kappa}(1) = su_2$. For $su_{1,1}$, $\kappa > 0$ corresponds to $2 k \kappa = 1$, 
where $k$ is the Bargmann index of the positive discrete series. For $su_{2}$, $\kappa < 0$ corresponds to $2 j \kappa = -1$, 
where $j$ is an angular momentum label associated with the ($2j$) irreducible representation of $su_{2}$. 

The $H$ Hamiltonian associated with ${\cal A}_{\kappa}(1)$ 
can be redefined as $H_0 = X_+ X_-$ with  		
	  \[
a^{\pm} = \frac{1}{\sqrt{b}}X_{\pm}, \quad 2 \kappa = \frac{a}{b}, \quad 
a \in \mathbb{R}, \quad b \in \mathbb{R}_+^* \Rightarrow 
H_0 = \frac{1}{2} a N (N-I) + bN  
	  \]		
and gives rise to three particular cases: 
(i)   $a = 0$, $b = 1$, $\kappa = 0$, $h_4$ case, $\dim {\cal F}_{\kappa} = \infty$: oscillator system; 
(ii)  $a = 1$, $2b = u + v + 1$, $u > 1$, $v> 1$, $\kappa > 0$, $su_{1,1}$ case, $\dim {\cal F}_{\kappa} = \infty$: P\"oschl-Teller system; and 
(iii) $a =-1$, $2b = \ell - 1$, $\ell \in \mathbb{N}\setminus\{0,1\}$, $\kappa < 0$, $su_2$ case, $\dim {\cal F}_{\kappa} = \ell$: Morse system. 

\section{Phase operators and phase states for ${\cal A}_{ \kappa }(1)$}
We now continue with the case where $r = 1$. We define the $E$ operator via 
	  \[ 
a^- = E \sqrt{F(N)} \quad {\rm with} \quad E \equiv E_{\infty} \quad {\rm or} \quad E \equiv E_{d}
	  \] 
for the infinite- or finite-dimensional case, respectively. 

\subsection{Case $\kappa \geq 0$: The $E_{\infty}$ operator}

The $E_{\infty}$ (phase) operator is not unitary 
($E_{\infty} E_{\infty}^{\dagger} = E_{\infty}^{\dagger} E_{\infty} + \vert 0 \rangle \langle 0 \vert = I$). Its 
eigenstates are phase states given by 
	  \[
E_{\infty} \vert \varphi, \theta \rangle = 
{\rm e}^{{\rm i} \theta} \vert \varphi, \theta \rangle, \quad 
\theta \in [-\pi , \pi], \quad 
\vert \varphi, \theta \rangle = \sum_{n=0}^{\infty} 
{\rm e}^{-{\rm i} F(n) \varphi}
{\rm e}^{{\rm i} n \theta} \vert n \rangle  
	  \]   
on $\{ \rho {\rm e}^{{\rm i} \theta} : \rho < 1 \}$. The $\vert \varphi, \theta \rangle$ phase states 
depend on two continuous parameters $\varphi$ and $\theta$, are neither normalized nor orthogonal, 
are temporally stable with respect to $H$, and satisfy a closure property ($\int_{-\pi}^{+\pi} d\theta 
\vert \varphi, \theta \rangle \langle \varphi, \theta \vert = 2 \pi  I$). 

The nonunitarity of $E_{\infty}$ requires a truncation procedure 
\textit{\`a la} Pegg and Barnett for the algebra $A_{ \kappa }(1)$ -- 
also useful for perturbation theory purposes. This leads to a truncated algebra ${\cal A}_{\kappa, s}(1)$ and a new representation space ${\cal F}_{\kappa, s}$ where $s$ is the order of truncation (the dimension of ${\cal F}_{\kappa, s}$). This situation is similar to the finite case corresponding to $\kappa < 0$ ($\dim {\cal F}_{\kappa} = d$) to be considered below. 

\subsection{Case $\kappa < 0$: The $E_d$ and $G_d$ operators}

\subsubsection{The $E_d$ operator}
The $E_{d}$ (phase) operator is unitary. Its eigenstates are 
given by 
	  \[
E_{d} \vert \varphi , m \rangle = 
{\rm e}^{2 \pi {\rm i} m/d} \vert \varphi , m \rangle, \quad 
\vert \varphi , m \rangle = \frac{1}{\sqrt{d}} \sum_{n = 0}^{d-1}
{\rm e}^{{\rm i}[-F(n) \varphi + 2 \pi mn/d]}
\vert n \rangle
    \]    
with $m \in \mathbb{Z}/{d\mathbb{Z}}$. The $\vert \varphi , m \rangle$ phase states depend on 
a discrete parameter $m$ and a continuous parameter $\varphi$; they are 
temporally stable with respect to $H$; and, for fixed $\varphi$, they are orthonormal 
($\langle \varphi , m \vert \varphi , m' \rangle =  \delta_{m,m'}$, 
the overlap $\langle \varphi , m \vert \varphi' , m' \rangle$ does not vanish for 
$\varphi' \not= \varphi$), they satisfy a closure relation 
($\sum_{m = 0}^{d-1} \vert \varphi , m \rangle \langle \varphi , m \vert = I$) and an equiprobability relation 
($\vert \langle n \vert \varphi , m \rangle \vert = 1/\sqrt{d}$). 

\subsubsection{The $G_{d}$ operator}
For $\kappa < 0$, we can define another type of phase operator, namely, 
	  \[
G_{d} = a^- + \frac{(a^+)^{d-1}}{F(d-1)!}, \quad F(n)! = F(1) F(2) \ldots F(n), 
\quad F(0) = 1. 
	  \]
The eigenstates of $G_{d}$ are phase states given by 
	  \[
G_{d} \vert \varphi , \mu \rangle = 
{\rm e}^{2 \pi {\rm i} \mu / d} \vert \varphi , \mu \rangle, \quad   
\vert \varphi , \mu \rangle = c_0 \sum_{n = 0}^{d-1} \frac{1}{\sqrt{F(n)!}}
{\rm e}^{{\rm i} [-F(n) \varphi + 2 \pi \mu n / d]} \vert n \rangle 
	  \]    
with $\mu \in \mathbb{Z}/{d\mathbb{Z}}$. The $\vert \varphi , \mu \rangle$ phase states depend on a discrete parameter $\mu$ and a continuous parameter $\varphi$, are temporally stable with respect to $H$, not normalizable, not orthogonal, and do not satisfy a closure property. The $\vert \varphi , \mu \rangle$ phase states are similar to the {Gazeau and Klauder coherent states except that: (i) they are eigenstates of a polynomial in $a^-$ and $a^+$, (ii) their labeling includes an integer, and (iii) they correspond to a finite spectrum (finite sum).

\section{Application to mutually unbiased bases}

By quantizing $\varphi$ according to $\varphi = - \pi (d-1) a / d$ with 
$a = 0, 1, \ldots, d-1$, the state vector $\vert \varphi , m \rangle$ becomes 
		\[
\vert \varphi , m \rangle \equiv \vert a m \rangle = \frac{1}{\sqrt{d}} \sum_{n = 0}^{d-1}
q^{n (d - n) a/2 + n m} \vert n \rangle, \quad 
q = \exp \left( \frac{2 \pi {\rm i}}{d}\right) 
		\]  
Note that $\vert a m \rangle$ 
is a quadratic discrete Fourier transform (the case $a = 0$ corresponds to an ordinary discrete Fourier transform). This quantization leads to the following result. For $d$ prime, the $d$ bases 
$\{ \vert a m \rangle : m = 0, 1, \ldots, d-1 \}$
for $a = 0, 1, \ldots, d-1$ and the canonical basis 
$\{ \vert n \rangle : n = 0, 1, \ldots, d-1 \}$
form a complete set of $d+1$ mutually unbiased bases (in the sense of quantum information). 

\section{Passing from ${\cal A}_{\kappa}(1)$ to ${\cal A}_{\kappa}(2)$}

As an extension of ${\cal A}_{\kappa}(1)$ (with one degree of freedom) 
we can define a one-parameter generalized Weyl-Heisenberg algebra 
${\cal A}_{ \kappa }(2)$ involving two degrees of freedom. The 
${\cal A}_{ \kappa }(2)$ algebra is spanned by $a_i^-$, $a_i^+$ and $N_i$ satisfying 
		\begin{eqnarray}
& & [a_i^- , a_i^+] = I + \kappa (N_1 + N_2 + N_i), \quad 
[a_i^{\pm} , a_j^{\pm}] = 0, \quad 
[N_i, a_j^{\pm}] = \pm \delta_{i,j} a_i^{\pm} 
		\nonumber \\
& & N_i = N_i^{\dagger}, \qquad 
a_i^+ = (a_i^-)^{\dagger}, \qquad 
[a_i^{\pm} , [a_i^{\pm} , a_j^{\mp} ]] = 0 \quad {\rm for} \quad i \not= j 
		\nonumber
		\end{eqnarray}
where $\kappa \in \mathbb{R}$ and $i, j \in \{ 1 , 2 \}$. This yields three cases: (i) $\kappa = 0 \rightarrow$ ${\cal A}_{\kappa}(2) = h_4 \otimes h_4$; (ii) $\kappa > 0 \rightarrow$ ${\cal A}_{\kappa}(2) = su_{2,1}$; and (iii) $\kappa < 0 \rightarrow$ ${\cal A}_{\kappa}(2) = su_3$.

In the infinite case ($\kappa \geq 0$), we can define two types of nonunitary phase operators (associated with $a_1^-$ and $a_2^-$). The corresponding phase states (neither normalized nor orthogonal) satisfy temporal stability and a closure property. A truncation {\it \`a la} Peggg and Barnett is necessary in view of the absence of unitarity. 

In the finite case ($\kappa < 0$), we can define four types of unitary phase operators. The phase states are here vector phase states (which are analogs of vector coherent states introduced by 
Hecht 1987 -- Zhang, Feng, Gilmore 1990 -- Thirulogasanthar, Ali 2003 -- Ali, Engli$\check{{\rm s}}$, Gazeau 2004 -- Ali, Bagarello 2005). They satisfy temporal stability, orthonormality, closure property, 
and equiprobability relation. They can be used for constructing mutually unbiased bases. 

\section{Coherent states for ${\cal A}_{ \{ \kappa \} }(1)$}

We come back to the case of a multi-parameter generalized Weyl-Heisenberg algebra ${\cal A}_{ \{ \kappa \} }(1)$. We shall deal in turn with coherent states of type I ({\it \`a la} Klauder and Perelomov) and of type II ({\it \`a la} Barut and Girardello). 

\subsection{Coherent states of type I}

The strategy is to look for states of the form 
$\vert z , \varphi \rangle = \sum_{n} {\overline{a_n}} z^n \vert n \rangle$ ($a_n \in \mathbb{C}$, $z \in \mathbb{C}$)
with the replacements $\vert n \rangle \rightarrow a_{n} z^{n}$ and 
$a^+ \rightarrow d/dz$. The derivation of the $a_n$ coefficients yields two results. 

In infinite dimension, the states
	  \[
	\vert z , \varphi \rangle = \sum_{n=0}^{\infty} 
  \frac{\sqrt{F(n)!}}{n!} z^n e^{-iF(n)\varphi} \vert n \rangle
		\]
exist only for $r=1$. They satisfy $\vert z , \varphi \rangle = \exp( z a^+) \vert 0 \rangle$
so that they are thus coherent states in the Klauder-Perelomov sense. 

In finite dimension ($dim = d$ or $s$), we have the same formal expression for $\vert z , \varphi \rangle$ except that
$\sum_{n=0}^{\infty} \to \sum_{n=0}^{dim-1}$ and $r = 1 \to r \ {\rm {arbitrary}}$. We thus get coherent states in 
finite dimension in the Klauder-Perelomov sense.

\subsection{Coherent states of type II}

The strategy here is to look for states of the form $\vert z , \varphi \rangle$ 
with the replacements $a_n \to b_n$, $\vert n \rangle \rightarrow b_{n} z^{n}$, 
and $a^+ \rightarrow {z}$. This leads to the two following results. 

In infinite dimension, the states
	  \[
\vert z , \varphi \rangle = \sum_{n=0}^{\infty} 
 \frac{1}{\sqrt{F(n)!}} z^n e^{-iF(n)\varphi} \vert n \rangle
		\]
exist for any value of $r$. They satisfy
$a^- \vert z , \varphi \rangle = z \vert z , \varphi \rangle$
and are thus coherent states in the Barut-Girardello sense. 

Note that in infinite dimension, the coherent states of types I and II 
are identical solely in the case where $\kappa_i = 0$ ($i = 1, 2, \cdots, r$). The corresponding states are the well-known Glauber states. 

In finite dimension ($dim = d$ or $s$), there are no Barut-Girardello coherent states for $z \in \mathbb{C}$. However, Barut-Girardello coherent states exist for
$z \to \theta = \ {\rm {Grassmann} \ {variable \ with}} \ \theta^{dim} = 0$. They are given by
	  \[
\vert \theta , \varphi \rangle = \sum_{n=0}^{dim-1} 
\frac{1}{\sqrt{F(n)!}} \theta^n e^{-iF(n)\varphi} \vert n \rangle
		\]	
and satisfy	
$a^- \vert \theta , \varphi \rangle = \theta \vert \theta , \varphi  \rangle$. 

The properties of the coherent states of type I and II are described in the papers mentioned in the abstract. 

\section{Closing remarks}

The developments in this paper should give an unified view of the three corresponding papers 
published by the authors. As open questions, let us mention the probabilistic interpretation 
and Bargmann functions associated with the coherent states of type I and II. 

\end{document}